\documentclass[aps,prc,superscriptaddress,twocolumn]{revtex4}
\usepackage{graphicx}
\usepackage{epsfig}%
\usepackage{bm}
\usepackage{hyperref}
\usepackage{color}

\begin{document}

\title{Impact parameter smearing effects on isospin sensitive observables in heavy ion collisions}

\author{Li Li}
\affiliation{China Institute of Atomic Energy, P. O. Box 275(10), Beijing 102413, China}
\affiliation{College of Physics and Energy, Shenzhen University, Shenzhen 518060, China}
\author{Yingxun Zhang}
\email{zhyx@ciae.ac.cn}
\affiliation{China Institute of Atomic Energy, P. O. Box 275(10), Beijing 102413, China}%
\affiliation{Guangxi Key Laboratory Breeding Base of Nuclear Physics and Technology, Guangxi Normal University, Guilin 541004, China}%
\author{Zhuxia Li}
\affiliation{China Institute of Atomic Energy, P. O. Box 275(10), Beijing 102413, China}
\author{Nan Wang}
\email{wangnan@szu.edu.cn}
\affiliation{College of Physics and Energy, Shenzhen University, Shenzhen 518060, China}%
\author{Ying Cui}
\affiliation{China Institute of Atomic Energy, P. O. Box 275(10), Beijing 102413, China}
\author{Jack Winkelbauer}
\affiliation{Physics Division, Los Alamos National Laboratory, Los Alamos, New Mexico 87545, USA }

\date{\today}

\begin{abstract}
The validity of impact parameter estimation from the multiplicity of charged particles at low-intermediate energies is checked within the framework of ImQMD model. The simulations show that the multiplicity of charged particles cannot estimate the impact parameter of heavy ion collisions very well, especially for central collisions at the beam energies lower than $\sim$70 MeV/u due to the large fluctuations of the multiplicity of charged particles.
The simulation results for the central collisions defined by the charged particle multiplicity are compared to those by using impact parameter b=2 fm and it shows that the charge distribution for $^{112}$Sn +$^{112}$Sn at 50 MeV/u is different evidently for two cases; and the chosen isospin sensitive observable, the coalescence invariant single neutron to proton yield ratio, reduces less than 15\% for neutron-rich systems $^{124,132}$Sn +$^{124}$Sn
at $E_{beam}$=50 MeV/u, while the coalescence invariant double neutron to proton yield ratio does not have obvious difference.
The sensitivity of the chosen isospin sensitive observables to effective mass splitting is studied for central collisions defined by the multiplicity of charged particles. Our results show that the sensitivity is enhanced for $^{132}$Sn+$^{124}$Sn relative to that for $^{124}$Sn+$^{124}$Sn, and this reaction system should be measured in future experiments to study the effective mass splitting by heavy ion collisions.
\end{abstract}


\pacs{Valid PACS appear here}
\maketitle


Knowledge about the isospin asymmetric nuclear equation of state (EoS), especially the density dependence of symmetry energy and neutron-proton effective mass splitting $m^*_{np}$ ($m^*_{np}=(m^*_n-m^*_p)/m$ or $f_I=\frac{1}{2\delta}(m/m_n^*-m/m^*_p)$) which reflects the momentum dependence of symmetry potential,
is of fundamental importance for our understanding of nature's most asymmetric objects including neutron stars and heavy nuclei \cite{Brown00, BALi08, Tsang12, Stein12,Horowitz14,Abbot17, Brueck55, BALi18}. Studies of nuclear structure, such as masses, neutron skins, and collective modes, have provided better knowledge about the EoS, symmetry energy and $m^*_{np}$ around normal nuclear density\cite{Horowitz14,Tsang12}. However, to study EoS for isospin asymmetric nuclear matter at densities far away from saturation density and at finite temperature, one can only rely on heavy ion collision experiments. By comparing the experimental data to transport model simulations, information about the EoS can be extracted. In order to obtain reliable physical information via the comparison of the experimental results with transport model simulations, it is essential that the simulated events must be in the same conditions, including beam energy, reaction geometry and other filters, as those of experiments.

In recent years, there has been a lot of effort on the extraction of symmetry energy and $m^*_{np}$ away from the normal nuclear density using data from heavy-ion collisions (HIC), by comparing measured isospin sensitive observables, such as isospin diffusion\cite{Tsang04,Sun10,Galichet09} at the beam energies from 35 to 74 MeV/u, double neutron to proton ratios\cite{Famiano06,Xie13,sujun16,Coupland16} at 50 and 120 MeV/u, angular distribution of neutron-excess for light charged particles at 35MeV/u\cite{RSWang14}, collective flows\cite{Kohley10,Russo16}, and pion ratios near the threshold energy\cite{Xiao09, BALi02, Song15, Tsang17} to transport model calculations. Consensus on the symmetry energy coefficient and slope of symmetry energy has been obtained from nuclear structure and reaction studies, where the symmetry energy at saturation density is $S_0=30-32$MeV and the slope of symmetry energy is $L=40-65$MeV\cite{Tsang12,Latti13}, even there is still large errors on these constraints.
Effective mass splitting has recently come up for debate, when HIC data was compared to transport models to extract the effective mass splitting in dense neutron-rich matter\cite{Coupland16,BALi18}. The comparison between the coalescence invariant double neutron to proton yield ratio (CI-DR(n/p)) data from National Superconducting Cyclotron Laboratory at Michigan State University (NSCL/MSU) at $E_{beam}$=120 MeV/u and ImQMD calculations favors the SLy4 interaction parameter set, which has $m^*_n<m^*_p$. Similar results have been observed in other transport codes calculations, such as Boltzmann-Langevin\cite{Xie13} and IQMD model \cite{sujun16}, even though they adopt the different forms of momentum dependent and density dependent of symmetry potential. This seems to conflict with the consensus that the sign of $m^*_{np}$ should be positive ($m_n^*>m^*_p$) from nuclear structure studies\cite{CXu10,XHLi15} and ab-initio calculations\cite{Bomba91,WZuo99,Hofm01}.
Furthermore, some BUU type codes\cite{Hermann14,Kong15} fail to reproduce these data, and IBUU calculations require explicit treatment of tensor and short-range correlations in reproduction of observed double ratios\cite{OHen15}.
Thus, effort is needed in the heavy-ion reaction community for understanding the effective mass splitting in dense neutron-rich matter\cite{Coupland16}. Two questions should be investigated. One is whether or not the sensitivity of CI-DR(n/p) to $m^*_{np}$ from $^{124}$Sn+$^{124}$Sn system is strong enough to develop useful constraints. The other is whether impact parameter smearing effect changes the results in the study of $m^*_{np}$ in simulations, as this effect has been noticed in the previous works\cite{Ogilvie89,Phair92} but many recent published papers ignore this effect\cite{Xie13,Zhang14,sujun16, Rizzo05,Colonna10,Kong15} in the study of isospin physics.

The impact parameter, which defines the entrance channel geometry, influences the heavy-ion collision (HIC) mechanism, such as energy or nucleon transfer for peripheral collisions and multifragmentation for more central reactions, and thus on it influences the observables, such as collective flows\cite{Tsang89PRC,Resid97}, charge distributions\cite{Zhang12}, isospin diffusion\cite{Zhang12,TXLiu07,Hudan03, Tsang04,Sun10,Galichet09}, angular distribution of emitted nucleons\cite{RSWang14, QHWu15,YZhang17} and its isospin contents. However, the impact parameter is not a direct measurable quantity, so it is usually estimated through the mean multiplicity of detected charged particles \cite{Ogilvie89, cavata90,Phair92, TXLiu12}, mass-weighted averaged parallel velocity of detected particles\cite{Peter90}, transverse kinetic energy of emitted light particles \cite{Lukasik97,Plagnol99}, flow angle\cite{Lecolley94}, nucleon multiplicity, longitudinal momentum transfer, quadrupole momentum tensor along beam direction $Q_{zz}$\cite{Tsang89}, or the combination of these methods using a neural network approach\cite{Bass96,David95}.
The impact parameter dependence of observables from BUU calculations in Ref.\cite{Tsang89} suggests that the mean multiplicity of fast nucleons and the linear momentum transferred to the target residue are relatively insensitive to the impact parameter for small impact parameters and incident energies below 60 MeV/u due to the fluctuations of the multiplicity measurement. Furthermore, large fluctuations of these observables for intermediate energy heavy-ion collisions cause the achievable accuracy of the impact parameter to be at best $\pm 0.2$ to $\pm0.3$ $b_{max}$ for central collisions\cite{Ogilvie89,Phair92,Peter90}. As a result, the impact parameter $b$ in simulations ranges from 0 to 0.4-0.6$b_{max}$ with the $b$ weighted gaussian-liked form\cite{Ogilvie89,Phair92,Peter90,Basrak16} for defined central collisions, rather than a simple form $b<0.3b_{max}$ or $b=0.2b_{max}$. This is known as impact parameter smearing. Thus, reliable constraints on the physics, such as the symmetry energy and effective mass splitting, through the comparison of HIC data to the transport model calculations, should be obtained by carefully considering this effect.


In this work, we firstly investigate the cause of impact parameter smearing and its influence on the accuracy of impact parameter estimation. The validity of the method of impact parameter estimation from the multiplicity of charged particles is discussed using ImQMD-Sky code\cite{Zhang14,Coupland16}.
Secondly, two isospin sensitive observables, the coalescence invariant single neutron to proton yield ratio (CI-R(n/p)) and the coalescence invariant double neutron to proton ratio (CI-DR(n/p)),
are analyzed by considering the impact parameter smearing effect in the simulations.
For understanding the sensitivity of CI-R(n/p) and CI-DR(n/p) to the effective mass splitting, two different Skyrme interaction parameter sets, SkM*\cite{Bartel} and SLy4\cite{Chaba97} are used. These two parameter sets have similar compressibility, symmetry energy coefficient and slope of symmetry energy but different values of isoscalar effective mass and effective mass splitting\cite{Zhang14}.


The central collisions we simulated are selected with the multiplicity of charged particles, which
is similar to that in the published CI-DR(n/p) data\cite{Coupland16,cavata90}.
The impact parameter is estimated from the following relation,
\begin{eqnarray}
\pi b_{est} (M \ge M_{0})&=&\pi b_{max}(\frac{\sum_{M_{0}}^{\infty}\sigma(M)}{\sum_{M_{min}}^{\infty}\sigma(M)})^{1/2},\\\nonumber
&=& \pi b_{max}(\frac{\sum_{M_{0}}^{\infty} N(M)}{\sum_{M_{min}}^{\infty} N(M)})^{1/2}.
\end{eqnarray}
Here, $M$ represents the multiplicity of charged particles, $\sigma(M)$ corresponds to the cross section with multiplicity equal to $M$, and $N(M)$ is the number of events with multiplicity equal to $M$. $b_{est}$ represents the estimated impact parameter with $M>M_0$, and $b_{max}$=1.12($A_{p}^{1/3}$+$A_{t}^{1/3}$). For given value of multiplicity ($M_0$), the estimated impact parameter is approximately determined as follows.
\begin{eqnarray}
\bar{b}_{est} (M_0)= \frac{1}{2}(b_{est} (M_{0}+\delta M)+b_{est} (M_{0}-\delta M)).
\end{eqnarray}
with $\delta M$=1 in this work.
The averaged real impact parameter $\bar{b}_{real}$ in ImQMD is defined as,
\begin{equation}
\bar{b}_{real}(M_{0})=\frac{\sum_{b}^{ } N(b,M_{0})b}{N(M_{0})}
\end{equation}
Here, N(\emph{b}, $M_{0}$) corresponds to the number of events which have the multiplicity $M_{0}$ at impact parameter \emph{b} in the simulations. The simulations are performed with event number $N_{total}$ from $b$=0 to $b_{max}$. The impact parameter $b$ for each event is determined from a Monte-Carlo sampling method within a circle with radius $b_{max}$, and $b=b_{max}\sqrt{\xi}$, where $\xi$ is a random number, between 0 and 1. By comparing the $\bar{b}_{real}$ and $\bar{b}_{est}$ at the same $M_0$, one can understand the accuracy of the impact parameter estimation method. From the N(\emph{b}, $M_{0}$), one can learn the smearing or variance of the impact parameter $b$ corresponding to a certain multiplicity $M_0$.

\begin{figure}[htbp]
\centering
\includegraphics[angle=0,scale=0.3]{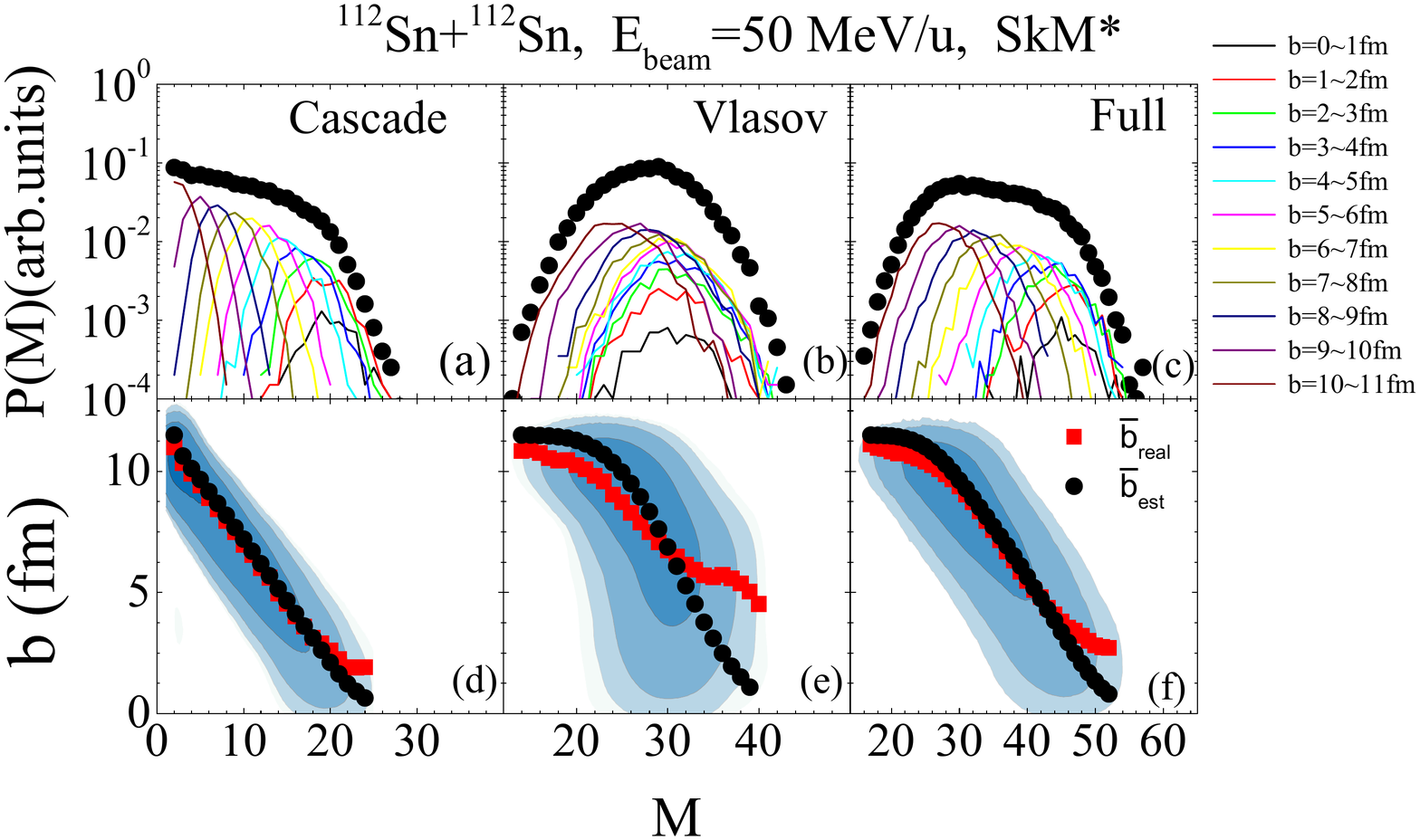}
\setlength{\abovecaptionskip}{0pt}
\caption{(Color online)
Upper panels show the integrated multiplicity distribution (black solid circle), as well as the multiplicity distribution for each impact parameter (colored lines).
Bottom panels show the $\bar{b}_{real}$ (red solid square) and $\bar{b}_{est}$ (black solid circle) as a function of M in ImQMD model, the contour plots are for event number at given b and M. (a) and (d) are for Cascade mode, (b) and (e) are for Vlasov mode, (c) and (f) are for Full mode. All the results are obtained with SkM* for $^{112}$Sn+$^{112}$Sn at $E_{beam}$=50 MeV/u.}
\setlength{\belowcaptionskip}{0pt}
\label{fig:fig1}
\end{figure}

\begin{figure}[htbp]
\centering
\includegraphics[angle=0,scale=0.32]{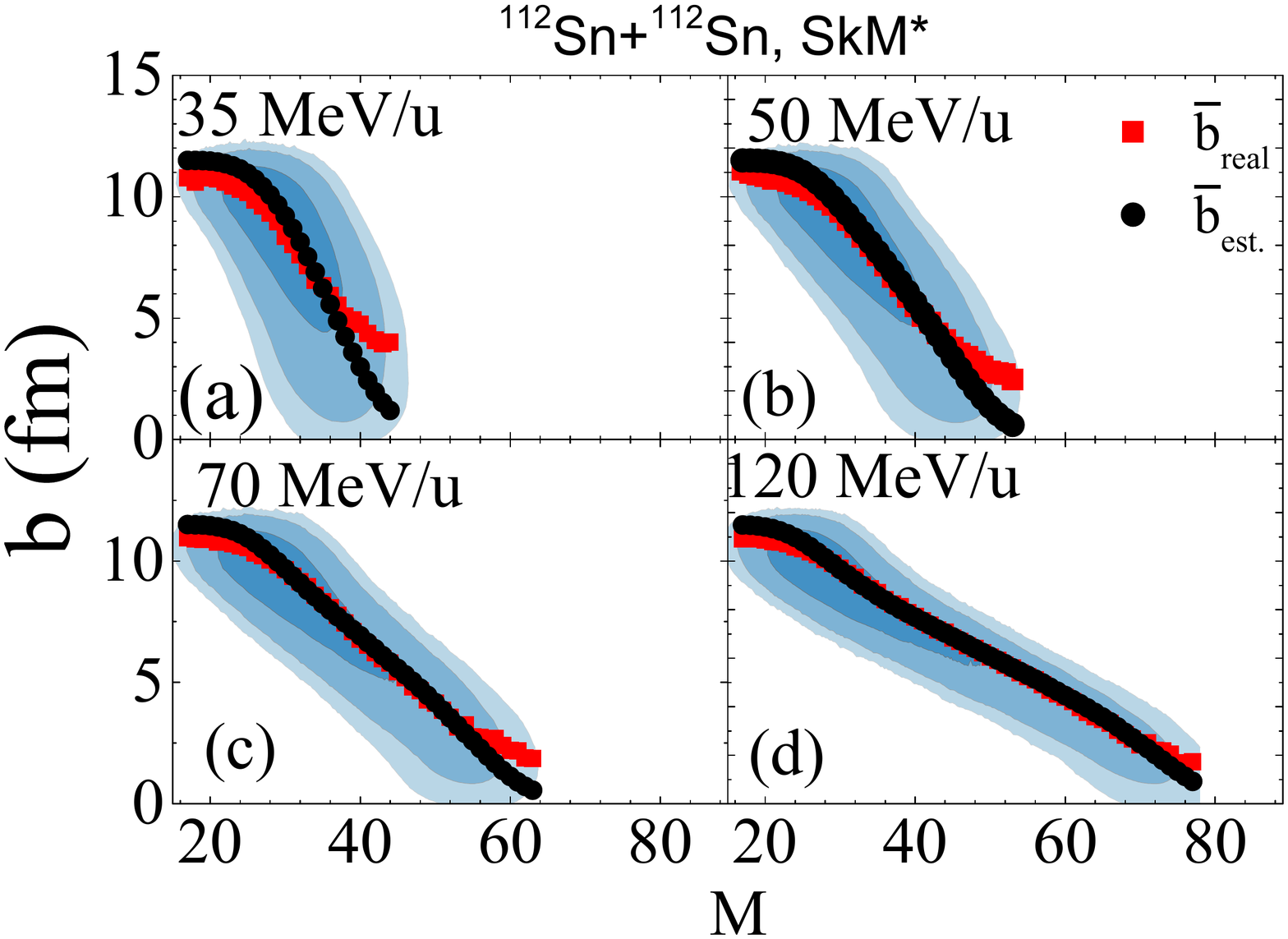}
\setlength{\abovecaptionskip}{0pt}
\caption{
(Color online)
Symbols are for $\bar{b}_{real}$ (red solid square) and $\bar{b}_{est}$ (black solid circle) as a function of M, the contour plots are for N(b,M). From left to right panels are for the beam energy of 35, 50, 70 and 120 MeV/u.}
\label{fig:fig2}
\setlength{\belowcaptionskip}{0pt}
\end{figure}

\begin{figure}[htbp]
\centering
\includegraphics[angle=0,scale=0.3]{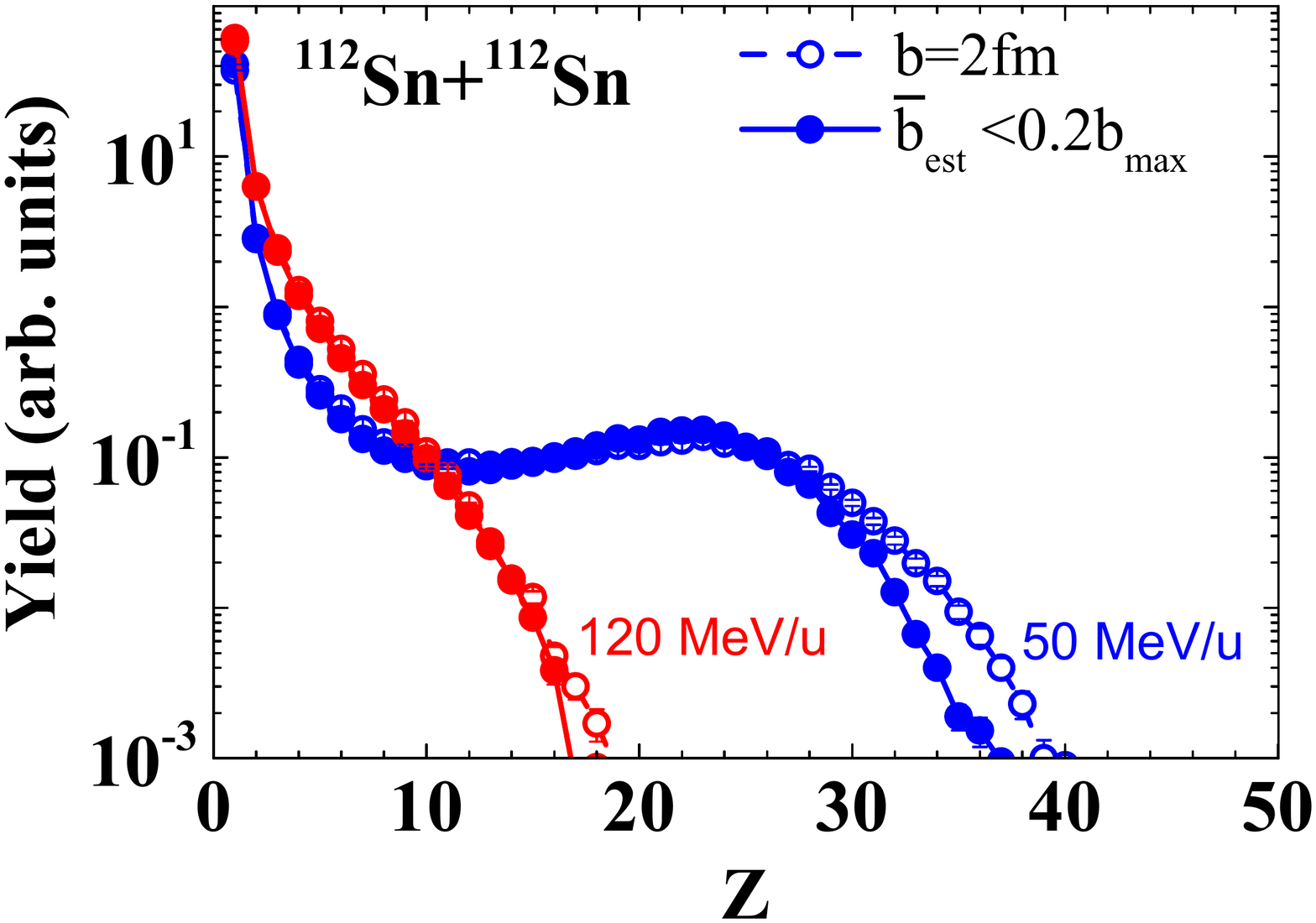}
\setlength{\abovecaptionskip}{0pt}
\caption{
(Color online) The charge distributions obtained with b=2fm (open circle) and $\bar{b}_{est}$ (solid circle), blue curves with symbols represent the results of $E_{beam}$=50 MeV/u and the red curves with symbols represent the results of $E_{beam}$=120 MeV/u. }
\label{fig:fig3}
\setlength{\belowcaptionskip}{0pt}
\end{figure}

In order to understand the cause of impact parameter smearing for low-intermediate energy heavy ion collisions, the effects of mean field potential and nucleon-nucleon collisions on the impact parameter estimation method are analyzed with ImQMD model. We investigated the difference between the averaged estimated impact parameter $\bar{b}_{est}$ and averaged real impact parameter $\bar{b}_{real}$ in three modes: 1) Cascade mode, i.e., only nucleon-nucleon collisions without mean field potential; 2) Vlasov mode, i.e., mean field potential without nucleon-nucleon collisions; and 3) Full mode, i.e., with both mean field potential and nucleon-nucleon collisions.
Figure~\ref{fig:fig1} (a) and (d) are the results for the Cascade mode; Figure~\ref{fig:fig1} (b) and (e) are for the Vlasov mode; Figure~\ref{fig:fig1} (c) and (f) are for the Full mode. The simulations are performed for $^{112}$Sn+$^{112}$Sn at the beam energy of 50 MeV/u, where the impact parameter $b$ is randomly chosen from 0 to $b_{max}$ as mentioned before. The simulated event number is 20,000 for generating Figure~\ref{fig:fig1} and ~\ref{fig:fig2}. In the Figures~\ref{fig:fig1}, ~\ref{fig:fig2} and ~\ref{fig:fig3}, the SkM* parameter set is used. The curves with different colors in upper panels of Figure~\ref{fig:fig1} represent the multiplicity distribution for different $b$ values, $b$=0-1, 1-2, ..., 10-11 fm. The black circle in the upper panels is the multiplicity distribution where $b$ is integrated over the whole range in the simulations.

For the cascade mode, we artificially set the initial Fermi momentum as zero to avoid the system breaking up before collision. This treatment reduces the nucleon-nucleon collision frequency compared to real heavy ion collisions, and makes the system more transparent. Consequently, the projectile and target residues tend to remain, and the multiplicity of fragments in this mode moves to the lower multiplicity region. The peak of the multiplicity distribution obviously moves to lower multiplicities as the impact parameter increases, except for $b<$1 fm. It results in the monotonic relationship between b and $<M>$. Thus, one can expect that $\bar{b}_{est}$ will agree well with $\bar{b}_{real}$ as in Ref.\cite{cavata90}, which is verified in Figure~\ref{fig:fig1} (d), where $\bar{b}_{est}$(black solid circle) agrees well with the results from averaged real impact parameter $\bar{b}_{real}$ (red solid square) in simulations.
Based on the results of N(b, M) in Figure~\ref{fig:fig1} (d), one can see the variance of impact parameter $b$ corresponding to certain M value is larger than 1 fm for cascade mode.

Figure~\ref{fig:fig1} (b) and (e) show the results calculated in Vlasov mode. The multiplicities are distributed over a larger region than that for cascade mode, while the peaks of the multiplicity distributions are located in a similar multiplicity region with similar shape for $b<$7 fm. At the energy studied in this work, the long range interaction plays a role and the overlap region of the system enters the  spinodal region during its evolution. The initial fluctuations in the QMD simulations leads to the large fluctuations in the observables, such as a broader multiplicity distribution. These behaviors result in the non-monotonic relationship between $b$ and the averaged values of multiplicity $<M>$, i.e., the mean field potential and fluctuations destroy the agreement between $\bar{b}_{est}$ and $\bar{b}_{real}$ for the whole impact parameter region, especially for $b<$7 fm. Due to the competition between the mean field potential and nucleon-nucleon collisions in the full simulation of heavy ion collisions, one can expect that the agreement between $\bar{b}_{est}$ and $\bar{b}_{real}$ should be between the results of Cascade and Vlasov. The results for the full mode are shown in (c) and (f), which shows that the impact parameter estimation method is good for $b>$4 fm for the beam energy of 50 MeV per nucleon, while the impact parameter varies significantly, about 2-3fm for a fixed multiplicity at this beam energy.

For a more global view on the validity of the impact parameter estimation, we investigate the reliability of $\bar{b}_{est}$ for the reaction $^{112}$Sn+$^{112}$Sn at beam energies ranging from 35 MeV/u to 120 MeV/u within the framework of ImQMD model. The results are shown in Figure~\ref{fig:fig2}. The $\bar{b}_{est}$ is close to $\bar{b}_{real}$ when $\bar{b}_{est}\ge$2 fm at the beam energy of 120 MeV/u. $\bar{b}_{est}$ starts to deviate from the $\bar{b}_{real}$ when $\bar{b}_{est}<$3, 4 and 6 fm at the beam energies of 70, 50 and 35 MeV/u, respectively. Based on the relation between $\bar{b}_{est}$ and $M$, the central, mid-peripheral and peripheral collisions in experiments can be sorted using the corresponding multiplicity region.
At the beam energy of 50 MeV/u, multiplicity region M$>$46\cite{Zhremark} in Figure~\ref{fig:fig2} corresponds to $\bar{b}_{est}/b_{max}$  $<$ 0.2 which is sorted as central collisions as in experiments. While the corresponding impact parameter in simulations ranges from 0 to 6 or 7 fm and the weights for b=1, 2, 3, 4, 5, 6 fm are 10-15\%, 22-26\%, 24-25\%, 21-22\%, 8-14\%, 2-6\%, $<$2\% (the exact values for the weight also depend slightly on the interaction parameter and size of reaction system we used in the calculations), respectively. This is consistent with the results in \cite{Ogilvie89,Phair92}. In the following text and figures, we use $\bar{b}_{est}/b_{max}$  $<$ 0.2 to represent the central collision defined by the multiplicity. The corresponding $M$ region and the weights for different $b$ at other beam energies are determined from the figure~\ref{fig:fig2}.

Now, let's turn to investigate the impact parameter smearing effect on HIC observables, such as the charge distribution, isospin sensitive observables, CI-R(n/p) and CI-DR(n/p) ratios for $^{112,124,132}$Sn+$^{112,124}$Sn. In Figure~\ref{fig:fig3}, we plot the charge distribution for $^{112}$Sn+$^{112}$Sn obtained with SkM*, with b=2 fm(dashed lines with open symbols), and $\bar{b}_{est}/b_{max}$ $<$ 0.2 corresponding to multiplicity region with M$>$46 (lines with solid symbols). The blue curves are the results for the beam energy at 50 MeV/u, and red curves are for 120 MeV/u. As we learned from Figure~\ref{fig:fig2}, the charge distribution obtained by the b=2fm and by the $\bar{b}_{est}/b_{max}<$0.2 (corresponding to M$>$70 in the ImQMD simulations) are very close at the beam energy of 120MeV/u, while the yields of heavy residues and intermediate mass fragments show disagreement at the beam energy of 50MeV/u. The charge distribution is narrower for the case of $\bar{b}_{est}/b_{max}<$0.2 than that for the b=2 fm case due to the selected events with high multiplicity for the $\bar{b}_{est}/b_{max}<$0.2 case. 

\begin{figure}[htbp]
\centering
\includegraphics[angle=0,scale=0.3]{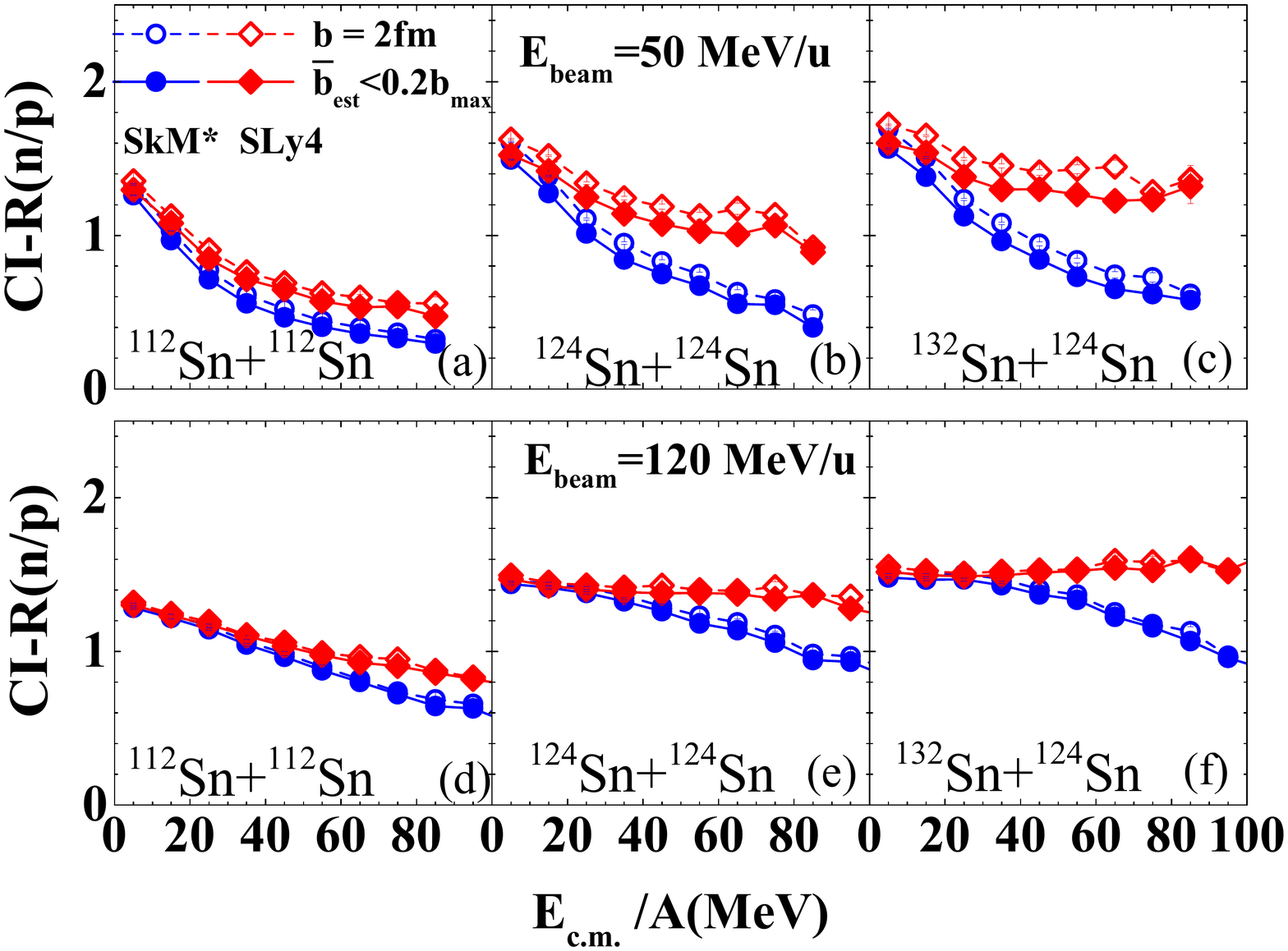}
\setlength{\abovecaptionskip}{0pt}
\caption{\label{fig:fig4}(Color online) The CI-R(n/p) ratios as a function of kinetic energy per nucleon, for $^{112}$Sn+$^{112}$Sn (left), $^{124}$Sn+$^{124}$Sn (middle), and $^{132}$Sn+$^{124}$Sn (right) obtained with SkM*(blue curve with circles) and SLy4 (red curve with diamonds). Open symbols are results obtained with b=2fm, and solid symbols are for $\bar{b}_{est}< 0.2 b_{max}$. Upper panels are the results for $E_{beam}$=50 MeV/u and bottom panels are for $E_{beam}$=120 MeV/u.}
\setlength{\belowcaptionskip}{0pt}
\end{figure}

For the isospin sensitive observables, coalescence invariant single neutron to proton yield ratios, i.e. CI-R(n/p)\cite{Famiano06, Zhang14}, and coalescence invariant double neutron to proton yield ratios, i.e. CI-DR(n/p), are studied. The CI-R(n/p) is defined as:
\begin{equation}
\mathrm{CI-R(n/p)}=\frac{Y^{CI}_{n}(E_{c.m.}/A)}{Y^{CI}_{p}(E_{c.m.}/A)}
\end{equation}
$Y^{CI}_{n,p}(E_{c.m.}/A)$ is the coalescence invariant spectral yield for neutron or proton, which is constructed from the combination of free nucleons and light fragments at same kinetic energy per nucleon as \cite{Famiano06, Zhang14},
\begin{eqnarray}\label{eqn:neu}
Y^{CI}_{p}(E_{c.m.}/A)&=&\sum Y(Z,A,E_{c.m.}/A) Z\\
Y^{CI}_{n}(E_{c.m.}/A)&=&\sum Y(Z,A,E_{c.m.}/A) (A-Z).
\end{eqnarray}
The Y(Z,A,$E_{c.m.}$/A) is the yield of nucleons or fragments (Z,A) with kinetic energy per nucleon $E_{c.m.}$/A, where the summation is up to A=6. The CI-DR(n/p) ratio is defined as,
\begin{equation}
\mathrm{CI-DR(n/p)}=\frac{\mathrm{CI-R_{a}(n/p)}}{\mathrm{CI-R_{b}(n/p)}}
\end{equation}
where $a=$ neutron-rich reaction system, and $b=$ neutron-poor reaction system. As in experiment\cite{Famiano06,Coupland16}, both the CI-R(n/p) and CI-DR(n/p) ratios are obtained within an angular gate of 70$^\circ$$<$$\theta_{c.m.}$$ <$ 110$^\circ$.
Calculations have verified that the CI-R(n/p) and CI-DR(n/p) at high kinetic energy still remain the same sensitivity to the density dependence of symmetry energy and effective mass splitting as the single n/p ratios for emitted free nucleons\cite{Rizzo05, Famiano06,Zhang14, Coupland16}. In Figure~\ref{fig:fig4}, we show the results for CI-R(n/p) obtained with the ImQMD-sky code for central collisions of $^{112,124,132}$Sn+$^{112,124}$Sn. The upper panels are the results for the beam energy of 50 MeV/u and lower panels show the results for 120 MeV/u. From the left to right the three reaction systems are shown, $^{112}$Sn+$^{112}$Sn, $^{124}$Sn+$^{124}$Sn, $^{132}$Sn+$^{124}$Sn, respectively.
The dashed lines with open symbols are the results obtained with b=2fm, and solid lines with solid symbols are with $\bar{b}_{est}/b_{max}<$0.2.
Compared to the results with b=2fm, the values of CI-R(n/p) obtained with $\bar{b}_{est}/b_{max}<$0.2 decrease by about 10 \% at high kinetic energy part for $^{124}$Sn+$^{124}$Sn and about 15\% for $^{132}$Sn+$^{124}$Sn at the beam energy of 50 MeV/u. This is because the central collision events defined by the particle multiplicity have a much higher yield of nucleons than that with b=2 fm, and this enhancement on the yield of nucleons slightly decreases the values of n/p due to the conservation of total nucleon number for reaction systems. At the beam energy of 120 MeV/u, the values obtained from b=2fm and $\bar{b}_{est}/b_{max}<$0.2 are almost the same, as we can expect from the comparison in Figure~\ref{fig:fig2}.

For investigating whether the sensitivity of the CI-R(n/p) to the effective mass splitting is changed due to the impact parameter smearing, two Skyrme interaction parameter sets, SkM* (blue curve with circles) and SLy4 (red curve with diamonds), are used in the following analysis.
Similar to the previous results for emitted free nucleons\cite{Zhang14}, the CI-R(n/p) obtained with SLy4 ($m_n^*<m_p^*$) is enhanced at high kinetic energy region, where the emitted nucleons are dominated by free nucleons and the strong Lane potential for $m_n^* < m_p^*$ enhances the neutron emission in neutron-rich reaction systems.
At lower kinetic energies, CI-R(n/p) becomes smaller relative to R(n/p) for free nucleons as in Ref.\cite{Zhang14}.
For the beam energy of 120 MeV/u, the sensitivity of CI-R(n/p) to effective mass splitting becomes weaker than for 50 MeV/u due to the increasing effects of the nucleon-nucleon scattering\cite{Zhang14}. The differences of CI-R(n/p) between SLy4 and SkM* increase with increasing isospin asymmetry of the reaction system for both beam energies.

The experimental CI-R(n/p) data has large uncertainties due to uncertainties in the efficiencies of neutron and charged particle detectors\cite{Famiano06}, which stimulates the remeasurement of coalescence invariant single n/p ratios for both 50 and 120 MeV/u for central collisions of $^{112,124}$Sn+$^{112,124}$Sn\cite{Tsang}. Because the data is not yet published, a comparison with the data is not made here. Nevertheless, the simulations at both 50 MeV/u and 120 MeV/u show that the sensitivity of CI-R(n/p) to the effective mass splitting becomes stronger for the very neutron-rich system, $^{132}$Sn+$^{124}$Sn, which should be measured in the further experiments by using a $^{132}$Sn beam\cite{Tsang17,Bisol}.

\begin{figure}[htbp]
\centering
\includegraphics[angle=0,scale=0.3]{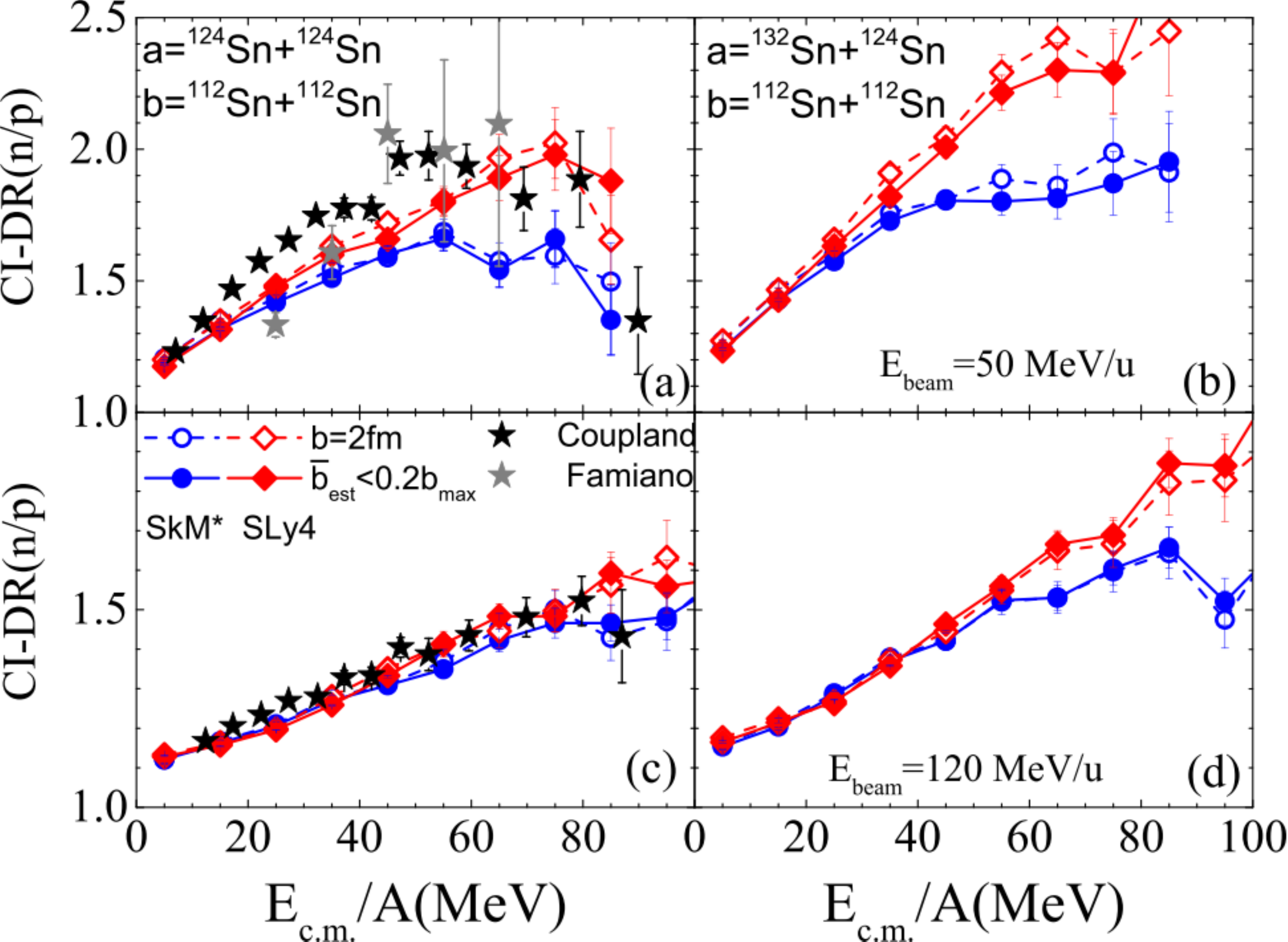}
\setlength{\abovecaptionskip}{0pt}
\caption{\label{fig:fig5} (Color online) CI-DR(n/p) as a function of kinetic energy per nucleon. Left panels are the results for a=$^{124}$Sn+$^{124}$Sn and right panels are for a=$^{132}$Sn+$^{124}$Sn, upperp panels are for 50 MeV/u and bottom panels are for 120 MeV/u. The legend of curves is similar to Figure~\ref{fig:fig4}.}
\setlength{\belowcaptionskip}{0pt}
\end{figure}

In Figure~\ref{fig:fig5}, we show the CI-DR(n/p) obtained with b=2 fm (dashed lines with open symbols) and $\bar{b}_{est}/b_{max}<$0.2 (solid lines with solid symbols) for SkM* (blue circles) and SLy4 (red squares). The left panels show a=$^{124}$Sn+$^{124}$Sn and b=$^{112}$Sn+$^{112}$Sn, and right panels show a=$^{132}$Sn+$^{124}$Sn and b=$^{112}$Sn+$^{112}$Sn. The results show that the CI-DR(n/p) obtained with $\bar{b}_{est}/b_{max}<$0.2 are very close to that obtained with b=2 fm.
The CI-DR(n/p) also depends on the effective mass splitting, but the sensitivity becomes weaker than that for CI-R(n/p).
Consistent with our previous results\cite{Famiano06, Coupland16}, the CI-DR(n/p) ratios calculated with $\bar{b}_{est}/b_{max}<$0.2 are below the data at the beam energy of 50 MeV/u, which may be related to the cluster formation mechanism in the transport models. At 120 MeV/u, the calculations with the SLy4 parameter set can reproduce the data better, but it hardly rules out the parameter set SkM* since the sensitivity of CI-DR to these two parameter sets is weak for $^{124}$Sn+$^{124}$Sn in the ImQMD simulations. This inspires us to use very neutron rich beams such as $^{132}$Sn. When changing the projectile from $^{124}$Sn to $^{132}$Sn, one finds the sensitivity of CI-DR(n/p) to the $m^*_{np}$ is enhanced compared that by using $^{124}$Sn projectile at both 50 MeV/u and 120 MeV/u, especially for high kinetic energy.

For the issue on the constraints of effective mass splitting by using HIC, one should also notice that the isoscalar effective masses for SkM* and SLy4 adopted in this work are different, where $m^*_s$=0.79$m$ for SkM* and $m^*_s$=0.69$m$ for SLy4. The different $m_s^*$ values could also influence the constraints on effective mass splitting\cite{Zhang15}. It has been found in the study of giant resonances using a dynamical approach\cite{Kong17}, the results show that the constrained sign of $m^*_{np}$ depends on the values of $m_s^*$. In these results, the GDR data supports $m_n^*<m_p^*$ when $m_s^*$=0.7$m$; while the data supports $m_n^*>m_p^*$ when $m_s^*$=0.84$m$. This result seems consistent with our comparison to the HIC data using the ImQMD model\cite{Tsang}. Nevertheless, reliable constraints on the effective mass splitting still need lot of work, such as using the statistical tools to do multi-variable analysis and combination of different isospin sensitive observables\cite{Tsang} to disentangle correlations among different physical parameters.

In summary, we have checked the accuracy of the impact parameter estimation from charged particle multiplicity at low-intermediate energy within the framework of ImQMD model. At the beam energies less than~70 MeV/u, large fluctuation causes the broader the multiplicity distribution, which destroys the monotonic correlations between the mean multiplicity of charged particles and the impact parameters, consistent with the results in\cite{Tsang89PRC}. For beam energies above 120 MeV/u where the nucleon-nucleon collisions play a more important role, the validity of this method becomes better.
By using charged particle multiplicity to specify central collisions at the beam energies we studied, the corresponding impact parameter $b$ in simulations should vary from zero to 5 or 6 fm for Sn+Sn collisions. These impact parameter smearing effects are taken into account in the simulations of heavy ion collisions.
Two isospin sensitive observables, CI-R(n/p) or CI-DR(n/p), for central collisions, are analyzed with the ImQMD model. The CI-R(n/p) ratios for central collisions obtained by using charged particle multiplicity decrease less than 15\% relative to collisions with b=2fm at the beam energy of 50 MeV/u for neutron-rich system, and the influence becomes smaller at the beam energy of 120 MeV/u. Furthermore, our studies also show that the sensitivity of CI-R(n/p) and CI-DR(n/p) to effective mass splitting can be enhanced for very neutron-rich reaction systems such as $^{132}$Sn+$^{124}$Sn. Experimental data with $^{132}$Sn+$^{124}$Sn would be very useful for understanding the effective mass splitting in the dense neutron-rich matter. 

\acknowledgments
This work has been supported by the National Natural Science Foundation of China under Grants No.(11475262, 11365004, 11375062, 11475115,11790320), National Key Basic Research Development Program of China under Grant No.(2013 CB834404, the High Precision Nuclear Physics Exeriments). Yingxun Zhang thanks useful comments by Prof. Jun Xu, Betty Tsang and Zhigang Xiao.


\begin{thebibliography}{99}
\bibitem{Zhremark} Due to the deficient on light particle formation mechanism in the transport models, the simulations always overpredict the yield of free nucleons. Consequently, the multiplicity for all charged particles is larger than the data at low-intermediate energy heavy ion collisions. The $\bar{b}_{est}$ as a function of M obtained in the model shift to larger M region relative to that from the data.
\end{thebibliography}
\end{document}